\documentclass{article}

\usepackage[left = 2cm, right = 2cm, top = 2cm, bottom = 2cm]{geometry}

\usepackage[comma, sort&compress]{natbib}

\usepackage[utf8]{inputenc}
\usepackage{authblk} 
\usepackage{setspace} 
\usepackage[english]{babel} 
\usepackage{mathpazo} 
\usepackage{amsmath, amsthm, mathtools, amsfonts} 
\usepackage{lineno} 
\usepackage[table]{xcolor} 
\usepackage{longtable, lscape} 
\usepackage[colorlinks=true,allcolors={blue!80!black}]{hyperref} 
\usepackage[capitalise, noabbrev]{cleveref} 
\usepackage{multirow} 
\usepackage[title,titletoc]{appendix} 

\usepackage{tikz}
\usetikzlibrary{matrix, calc, shapes.geometric, arrows}
\tikzstyle{outcome} = [rectangle, rounded corners, minimum width = 3cm, minimum height = 1cm, text centered, text width = 3cm, draw = black]
\tikzstyle{reason} = [rectangle, rounded corners, minimum width = 10cm, minimum height = 0.75cm, text width = 10cm, draw = black]
\tikzstyle{arrow} = [thick, ->, >= stealth]

\usepackage{caption} 
\usepackage{subcaption} 
\usepackage{graphicx} 
\usepackage{rotating} 

\makeatletter
\def\hlinewd#1{%
  \noalign{\ifnum0=`}\fi\hrule \@height #1 \futurelet
   \reserved@a\@xhline}
\makeatother

\newcommand{\inla}{\texttt{r-inla}}

\newcommand{\pa}[1]{\left(#1\right)} 
\newcommand{\rOpenInt}[1]{\left[#1\right)} 
\newcommand{\logit}{\text{logit}} 
\newcommand{\bhf}{\text{Bayesian hierarchical framework}} %

\graphicspath{ {./figures/} }

\defcitealias{ukhls2019weights}{UKHLS, 2019}
\defcitealias{ukhls2022data}{UKHLS, 2022}  
\defcitealias{dwp2022statxplore}{DWP, 2022a} 
\defcitealias{dwp2022ucStatistics}{DWP, 2022b} 
\defcitealias{dwp2022ucAge}{DWP, 2022c} 
\defcitealias{ons2019imd}{ONS, 2019} 
\defcitealias{ons2021census}{ONS, 2021} 
\defcitealias{ons2022bame}{ONS, 2022a} 
\defcitealias{ons2022employmentEthnicity}{ONS, 2022b}
\defcitealias{ons2023geog}{ONS, 2023}
\defcitealias{ifs2020catching}{IFS, 2020} 
\defcitealias{hes2017well}{HES, 2017} 

\title{Bayesian Interrupted Time Series for evaluating policy change on mental well-being: an application to England's welfare reform}

\author[1*]{C. Gascoigne}
\author[1]{M. Blangiardo}
\author[2]{Z. Shao}
\author[3]{A. Jeffery}
\author[4]{S. Geneletti}
\author[3]{J. B. Kirkbride}
\author[2]{G. Baio}

\affil[1]{MRC Centre for Environment and Health, Department of Epidemiology and Biostatistics, School of Medicine, Imperial College London, London, UK}
\affil[2]{Department of Statistical Science, University College London, London, UK}
\affil[3]{Division of Psychiatry, University College London, Psylife Group, London, UK}
\affil[4]{Department of Statistics, London School of Economics and Political Science, London, UK}
\affil[*]{Corresponding author. Email: c.gascoigne@imperial.ac.uk}

\date{}

\begin{document}

\maketitle

\abstract{Factors contributing to social inequalities are also associated with negative mental health outcomes leading to disparities in mental well-being. We propose a Bayesian hierarchical model which can evaluate the impact of policies on population well-being, accounting for spatial/temporal dependencies. Building on an interrupted time series framework, our approach can evaluate how different profiles of individuals are affected in different ways, whilst accounting for their uncertainty. We apply the framework to assess the impact of the United Kingdom’s welfare reform, which took place throughout the 2010s, on mental well-being using data from the UK Household Longitudinal Study. The additional depth of knowledge is essential for effective evaluation of current policy and implementation of future policy.}

\section{Introduction}

Factors contributing to social inequalities such as low income, unemployment and poorer education are strongly associated with worse mental health \citep{marmot2020health} and, in particular depression, anxiety \citep{muntaner2004socioeconomic}, psychosis \citep{o2016neighbourhood} and, self-harm and suicide \citep{lorant2018socioeconomic}. Several studies have shown that individuals exposed to the greatest socioeconomic disadvantage, including poverty, are more likely to experience mental health problems \citep{singh2019housing, rose2020social, byrne2020ethnicity} which heightens disparities in mental well-being related illness across socioeconomic status. Changes in policy that directly involve those who are considered more socially disadvantaged can have a profound effect on their mental well-being. Therefore, it is important to ensure any changes to these policies do not have a negative effect in general, and if there is a negative effect, do not exacerbate the already disproportionate effect on those most at risk. 

Evaluating the causal effects of such policies on mental health outcomes in the population should be used to inform whether and how local, regional, and national policies should be implemented and modified. However, studies and, flexible, generalisable frameworks to quantify these effects are rarely employed for this purpose. In this context, to evaluate the impact of policies, randomised controlled trials are not viable and observational studies are usually employed instead. Despite their practical utility, working with observational (quasi-experimental) designs has potential problems \citep{campbell2015experimental, cook2002experimental}, e.g., the possible presence of pre- and post-intervention trends, as well as the need to find comparable control groups to deal with residual confounding, which might occur if, for instance, variables related to the outcome change rapidly (e.g., some disease outbreaks) \citep{linden2011applying, abadie2003economic, bernal2017interrupted}.


An example of a recently implemented policy reform that directly impacts those of a lower socioeconomic status in the United Kingdom (UK) is the welfare reform known as Universal Credit (UC). UC was introduced in the early 2010s as part of a process of welfare reforms initiated by the coalition government led by then Prime Minister David Cameron of the Conservative Party, to replace six separate welfare benefits (i.e., Child Tax Credit, Housing Benefit, Income Support, income-based Jobseeker’s Allowance, income-related Employment and Support Allowance, and Working Tax Credit) by a single unified welfare benefit \citep{ukparliment2020aims}. The process was intended to be simpler and facilitate access and receipt of welfare according to need. However, the implementation of UC was marred by controversies, including a lengthy delay in payment and increased sanctions, meaning that individuals received reduced amounts or no amount of welfare support at all, sometimes for prolonged periods \citep{craig2020early, cheetham2022exploring, mahase2020universal}. To the best of our knowledge, there has only been one longitudinal study that evaluated the early relationship between Universal Credit and mental well-being \citep{wickham2020effects}. The authors used longitudinal panel data from the United Kingdom Household Longitudinal Study (UKHLS, or ``Understanding Society'') and showed that when UC was `introduced' to a local authority (i.e., at least one person began receiving UC), those who were unemployed suffered a disproportionate decline in self-reported mental health. 

The analysis performed by \citet{wickham2020effects} provided important insights into the difference between employed and unemployed participants whilst considering several influential demographic confounders such as age, education, marital status, and sex. However, in addition to a limited follow up, it did not account for any differences due to geographical (spatial) location. Spatial location is important in scenarios such as these as geography is often a surrogate for any residual confounding not already captured; consequently, it can be seen as a proxy for social, health, and environmental variables. For example, location can be important for quality of health services \citep{corris2020health, ellis2010regional}, and wealth and economy \citep{forth2021regional, marmot2020health, ifs2020catching}. Differences due to location can obscure the true effect of the government policy. Therefore, capturing any residual variation caused by geographical location is vital when evaluating the effect of government policy.

In this paper, we propose a flexible hierarchical Bayesian model to examine the causal effects of policy change on mental health outcomes. We account for both spatial and (non-linear) temporal trends, as well as evaluate inequalities indexed by deprivation and ethnicity. In contrast to \citet{wickham2020effects}, in our analysis (framed in an \textit{Interrupted Time Series} perspective \citep{wagner2002segmented, penfold2013use}), we explore the relationship between self-reported mental well-being and a so-called `contextual awareness' to UC. This is defined at the local authority level and is a measure of how many people are aware of the UC roll-out in a particular area. This accounts for the fact that even those still not in the scheme are likely to be aware of (and potentially influenced by) it. We examined how sensitive the effect estimate is to different definitions of contextual awareness. We hypothesised that contextual awareness of UC would lead to worse self-reported psychological distress amongst those exposed to it relative to those who were unexposed.

The remainder of the paper is organised as follows: in Section 2, we outline the selected participants, formally define the measured outcome, the exposed and control population, the definition of contextual awareness, any associated confounding variables, the statistical model, a measure the effect of the intervention, and describe the implementation strategy; in Section 3, we include the results from a sensitivity analysis to the definition of a contextual awareness, and the results broken-down into different temporal, spatial and socioeconomic profiles; and Section 4 concludes the paper with a discussion.

\section{Methods} \label{Section: Method}

We analysed individual-level mental health outcomes as yearly self-reported responses about psychological distress from Understanding Society \citepalias{ukhls2022data}. The UKHLS dataset is a representative longitudinal panel survey based on a stratified cluster sampling design. The sample consists of approximately 40,000 households that were first interviewed in 2009 and have been followed since in waves that span roughly three years. Specifically, we have data on years 2009--2021. Interviews are either conducted face-to-face by trained individuals or can be completed online. The purpose of the UKHLS is to provide insight into a range of different topics that include work, education, income, family, social life, and health (including mental well-being).

\subsection{Selected participants}

We included individuals who were of a working age (16--64 years old \citepalias{dwp2022ucAge}) and had information (i) on employment status, (ii) Lower Tier Local Authority of residence (LTLA; a total of 317 administrative geographies in England defined by the Office for National Statistics for the purpose of local government \citep{gov2016ltla}), (iii) the outcome (psychological distress) and (iv) confounding variables (discussed below). We excluded all individuals who did not reside within England (i.e., Scotland, Wales, and Northern Ireland) to combine the UKHLS data with the 2019 Index of Multiple Deprivation \citepalias[IMD;][]{ons2019imd}, an established measure of relative social deprivation for small areas. We excluded people who reported life-time sickness or a disability as they would not qualify for UC, while at the same time be more likely to experience mental ill-health \citep{wickham2020effects}.

After the selection process, we had 380\,378 observations from 47\,555 distinct individuals followed from 2009--2021 from the UKHLS survey data. Of those individuals, the average age was 40, there was 53.80\% female, 27.40\% of the non-white ethnicity, and 15.30\% were unemployed at some time during the study period.

\subsection{Measured outcomes}

We derived psychological distress using self-reported answers to the General Health Questionnaire-12 (GHQ-12), a series of 12 non-leading questions that gives an insight into an individual’s general psychological distress over the previous few weeks \citep{gnambs2018structure}. An example question is ``Have you been able to manage your problems in the past few weeks?’’ The respondents answer each question on a scale of 0--3 mapping onto ``non-distress'' to ``distress''. Following the recommendation of the Health Survey for England \citepalias{hes2017well}, we summed these scores into a value ranging between 0 and 12, and then dichotomised it as follows: scores of 0--3 indicated no psychological distress, and scores of 4--12 indicated psychological distress.

\subsection{Exposure}

For each individual, we defined the exposure based on their yearly self-reported answers to the employment status question in the UKHLS survey. Specifically, the exposure of interest is UC. However, since the UKHLS does not have sufficient information on whether an individual is receiving UC, we use each respondent's employment status as a proxy for exposure to UC. The respondents were asked ``Which of these best describes your current employment situation?'' and were given 14 options. Given our selection criteria, we take all those of a working age who responded ``unemployed'' as the exposed population, and all other responses, excluding ``life-time sick or disabled'' as the control population.

\subsection{Intervention - contextual awareness to Universal Credit}

Due to the roll-out nature of UC (different LTLAs adopt the new policy at separate times, and the transition of individuals from the previous welfare system to UC was at different rates), there is no clear definition of when the intervention has begun. Hence, we defined the intervention to have begun in a specific LTLA when the LTLA becomes ``contextually aware'' of UC. 

Let us consider an individual living in a given LTLA who is receiving the legacy welfare benefit but is aware of others within their LTLA who have already transitioned onto UC. Whilst they might not be on UC themselves, the anticipation of their impending transition has the potential to cause them psychological distress as they become aware of the issues surrounding UC (well documented lengthy delays and increased sanctions as reported by numerous media outlets). Whilst we cannot be sure of an individual’s awareness to UC, we stipulate that if enough people within the LTLA are on UC (25\% of the number of individuals on UC in the study period's last monthly total), then the LTLA is defined to have become ``contextually aware'' of UC and the intervention has begun. 

To obtain a measure of contextual awareness of UC we used monthly statistics from the Department of Work and Pensions \citepalias[DWP;][]{dwp2022statxplore} based on monthly totals of the number of people in each LTLA registered to receive UC. Specifically, we use the first time an LTLA passed the threshold of 25\% of the December 2021 monthly total (i.e., the last month of observation in our study). The choice of threshold percentage was arbitrary, and the definition of the intervention beginning was dependent upon this. Consequently, we performed a sensitivity analysis to see how robust the results are to difference in changes to the definition of contextual awareness. We compared the results using 5\%, 15\%, 25\%, 35\% and 45\% as the percentages. In a second sensitivity analysis, we included the results for when an LTLA was first introduced to UC, as in \citet{wickham2020effects}.

\subsection{Confounders}

As individual-level confounders, we considered each participant's age, education level, ethnicity, marital status, and sex. These are common choices to account for demographic and socioeconomic confounding. We considered ethnicity as a particularly important confounder because it is related to both employment status and level of psychological distress \citep{barnes2017racial, zuccotti2019ethnicity}. We also included two area-level confounders, in order to capture potential residual confounding at the LTLA level: (i) social deprivation, as measured by the 2019 IMD \citepalias{ons2019imd}, and (ii) ethnic mix defined as the proportion of the population from Black, Asian, or other minority ethnic groups as reported in the 2021 census \citepalias{ons2022bame}. The area-level confounders were defined at the Lower Layer Super Output Areas (LSOAs), an administration level comprising of 400--1200 individuals with over 33\,000 in England, designed to improve the reporting of small area statistics \citep{ons2021census}. We grouped these based on deprivation and ethnic mix characteristics. For deprivation, we ranked each LSOA based on their IMD score and grouped them into deciles from 1 (most deprived) to 10 (least deprived). For ethnic mix, we grouped the LSOAs into quintiles (20\% intervals) defined by the proportion of the population from ethnic minorities (excluding white minorities) from 1 (most ethnically mixed) to 5 (least ethnically mixed).

\subsection{Statistical model}

Psychological distress, considered as binary, was modelled using a Bernoulli random variable, for each individual ($i = 1, \dots I_l$) in each LTLA ($l = 1, \dots, 309$) and each year ($t=1,\ldots,T$):  
\begin{equation*}
    y_{itl} \sim \text{Bernoulli}\pa{p_{itl}}
\end{equation*}
where $p_{itl}$ is the underlying probability. Using the standard logit transform, the Interrupted Time Series (ITS) model with controls can be written as
\begin{equation}
    \label{Eq: BITS model}
    \begin{split}
        \text{logit} & \pa{p_{itl}} = \mu_{itl} = \beta_0 + \texttt{year}_{l} \beta_1 + \texttt{intervention}_{lt} \beta_2 + \texttt{year}^{+}_l \beta_3 +\\
        & \texttt{exposed}_{it}\beta_4 + (\texttt{year}_l \times \texttt{exposed}_{it}) \beta_5 +\\ 
        & \pa{\texttt{intervention}_{lt} \times \texttt{exposed}_{it}}  \beta_6 +
         \pa{\texttt{year}^{+}_l \times \texttt{exposed}_{it}}  \beta_7  +\\
        & \pa{\texttt{age}_{it}}  \beta_8 + \pa{\texttt{education}_{it}}  \beta_9 +\\
        & \pa{\texttt{ethnicity}_{i}}  \beta_{10} + \pa{\texttt{marital status}_{it}}  \beta_{11} + \pa{\texttt{sex}_{i}}  \beta_{12} +\\
        & \pa{\texttt{deprivation}_{l}}  \beta_{13} + \pa{\texttt{ethnic mix}_{l}}  \beta_{14} +\\
        & \gamma_t + \delta_l,
    \end{split}
\end{equation}
where $\texttt{intervention}$ is a binary variable which indicates whether the LTLA has become contextually aware of UC at time $t$,  $\texttt{year}$ is a discrete variable, centered on the year of the UC awareness for each LTLA, and $\texttt{year}^{+}$ is a discrete variable indicating the number of years since the intervention occurred for each LTLA.

A typical ITS model includes an intercept, a linear trend in time before the intervention (i.e., $\texttt{year}$), the immediate effect of the intervention (i.e., $\texttt{intervention}$), and a linear trend in time after the intervention (i.e., $\texttt{year}^{+}$) only \citep{bernal2017interrupted}. Here, to account for general trends in psychological distress over the study period, we include an \texttt{exposure} variable, which distinguishes exposed individuals from the control ones, as well as allowing for interactions between the ITS terms and the exposure group. In particular, in Equation \ref{Eq: BITS model}, the ITS terms are $\left\{\beta_0, \beta_1, \beta_2, \beta_3\right\}$ for the control group, and $\left\{(\beta_0+\beta_4), (\beta_1+\beta_5), (\beta_2+\beta_6), (\beta_3 +\beta_7)\right\}$ for the exposure one, respectively. The additional regression coefficients in the model $\left\{\beta_8 \ldots \beta_{14}\right\}$ are for the individual- and LTLA-level confounders. The final two terms, $\gamma_t \sim \text{Normal}\pa{0, \sigma_\gamma^2}$ and $\delta_l \sim \text{Normal}\pa{0, \sigma_\delta^2}$, are unstructured random effects used to account for any residual variation in time and space, respectively, not captured by the fixed effects. The random effects provide global smoothing across the time and spatial field of the study independently of one another. 

For the intercept, we used a non-informative, uniform prior $\beta_0 \sim \text{Uniform}\pa{-\infty, +\infty}$. For the fixed effects we used weakly-informative, normal prior $\beta_1, \dots, \beta_{14} \sim \text{Normal}\pa{0, 1000}$. For the standard deviations in the temporal and spatial random effects, we used a penalised complexity prior $P\pa{\sigma_\gamma > 1} = 0.1$ and $P\pa{\sigma_\delta > 1} = 0.1$, respectively. 

\subsection{Survey weights}

The UKHLS provides two sets of weights (cross-sectional and longitudinal) for any analysis that wishes to account for the survey design. The cross-sectional weights are for analyses that use one waves worth of data, while the longitudinal weights are for those studies using multiple waves. A drawback of the longitudinal weights is that they only consider those individuals who have responded in all waves. To include individuals who may have missed one or more interviews, we use the method recommended by the UKHLS \citepalias{ukhls2019weights}. Briefly, we take the cross-sectional weights from the first wave and adjusts these based on the individual’s probability of responding to all subsequent time points. The clear advantage is that we do not disregard individuals who miss a wave; however, we can only capture those present in the first wave. Similar adjustments have been made in other work on mental health outcomes from the UKHLS \citep{dotsikas2023trajectories}.

\subsection{Standardised percentage change}

To describe the overall impact of UC on mental well-being on the exposed population, we constructed a standardised (percentage) change, which estimated the change in distress for the exposed population before and after they experienced contextual awareness of UC, adjusting for the same difference in the controls. Dropping all the indices for simplicity, let E $ = $ Exposed, C $ = $ Controls, A $ = $ After, and B $ = $ Before, where A and B are for all the years in the study period after and before the intervention, respectively. The prevalence of psychological distress in the exposed population after experiencing contextual awareness to UC is defined as
\begin{equation*}
    \label{Eq: psychological distress prevalence in the exposed after UC}
    p^\text{EA} = \logit^{-1}\left(\mu^{^\text{EA}}\right),
\end{equation*}
where the superscript EA indicates the combination of regression terms in \ref{Eq: BITS model} for the exposed population after the intervention has been implemented. Similarly, the prevalence of psychological distress in the exposed population before the intervention is defined as   
\begin{equation*}
    \label{Eq: adjusted psychological distress prevalence in the exposed before UC}
    \tilde{p}^\text{EB} = \logit^{-1}\left(\mu^\text{EB} \frac{\mu^\text{CA}}{\mu^\text{CB}}\right).
\end{equation*}
where we include the term  $\frac{\mu^\text{CA}}{\mu^\text{CB}}$ to adjust for anything that might have impacted the outcome at population level in the period after vs before the implementation of the policy.
Hence we used  the notation $\tilde{p}$ to indicate that this is a standardised prevalence. If $\frac{\mu^\text{CA}}{\mu^\text{CB}}$ is above 1, the prevalence before for the exposed groups will be inflated to account for the fact that a change before-after is also seen in the control group, hence the effect of the policy implementation will be smaller than if considering directly the difference between before-after in the exposed population. A similar approach was used for evaluating health risks from the opening of municipal waste incinerators \citep{freni2019bayesian}. Finally, we define
\begin{equation}
    \label{Eq: Standardised change}
    \rho = \pa{p^\text{EA} - \tilde{p}^\text{EB}} / {\tilde{p}^\text{EB}}
\end{equation}
as the standardised change, where $\rho > 0$ indicates an increase in the prevalence of psychological distress following UC.

\subsection{Implementation}

We fitted the ITS model within a Bayesian hierarchical framework using Integrated Nested Laplace Approximation (INLA), as implemented in the \inla \ package \citep{rue2009approximate}. INLA provides accurate approximations of the marginal posterior distribution for all model parameters whilst avoiding the need for costly and time-consuming Markov-chain Monte Carlo (MCMC) sampling. Unlike an MCMC implementation, which samples directly from the joint posterior distributions, INLA produces approximations to the marginal distributions of the model parameters. These are then post-processed into additional quantities of interests, such as joint posterior distribution.

For individual $i$, time $t$ and LTLA $l$, the $\pa{n}^{th}$ draw from the posterior distribution of the linear predictor is $\mu^{\pa{n}}_{itl}$. For a given draw of the linear predictor’s posterior distribution, we can define a range of additional marginal posterior distributions. For example, aggregating draws over specific time(s), location(s), confounder(s), considered alone or in combination, we can obtain a marginal distribution for areas, specific times and covariate profiles. For instance, marginalising the draws from the posterior distribution for the exposure group  before-after the intervention, we obtain draws from the posterior distribution of $\mu^{\text{EB}, \pa{n}}$ and $\mu^{\text{EA}, \pa{n}}$. Similarly we can obtain $\mu^{\text{CB}, \pa{n}}$ and $\mu^{\text{CA}, \pa{n}}$ from the draws related to the controls and in turn define a draw from the posterior distribution of the standardised change $\rho^{\pa{n}}$. In the following results section, we explore the range of termed profiles and for each, presented the posterior medians and 95\% Credible Interval (CrI).

\section{Results}

\subsection{Exposure groups by temporal profile}

\begin{figure}[!h]
    \centering
    \caption{National average of psychological distress for exposed (red) and control (blue) populations. The solid line represents the median value of the posterior distribution, and the shaded region is the 95\% Credible Interval.}
    \label{Fig: centeredPlot.national}
    \includegraphics[width=0.5\textwidth]{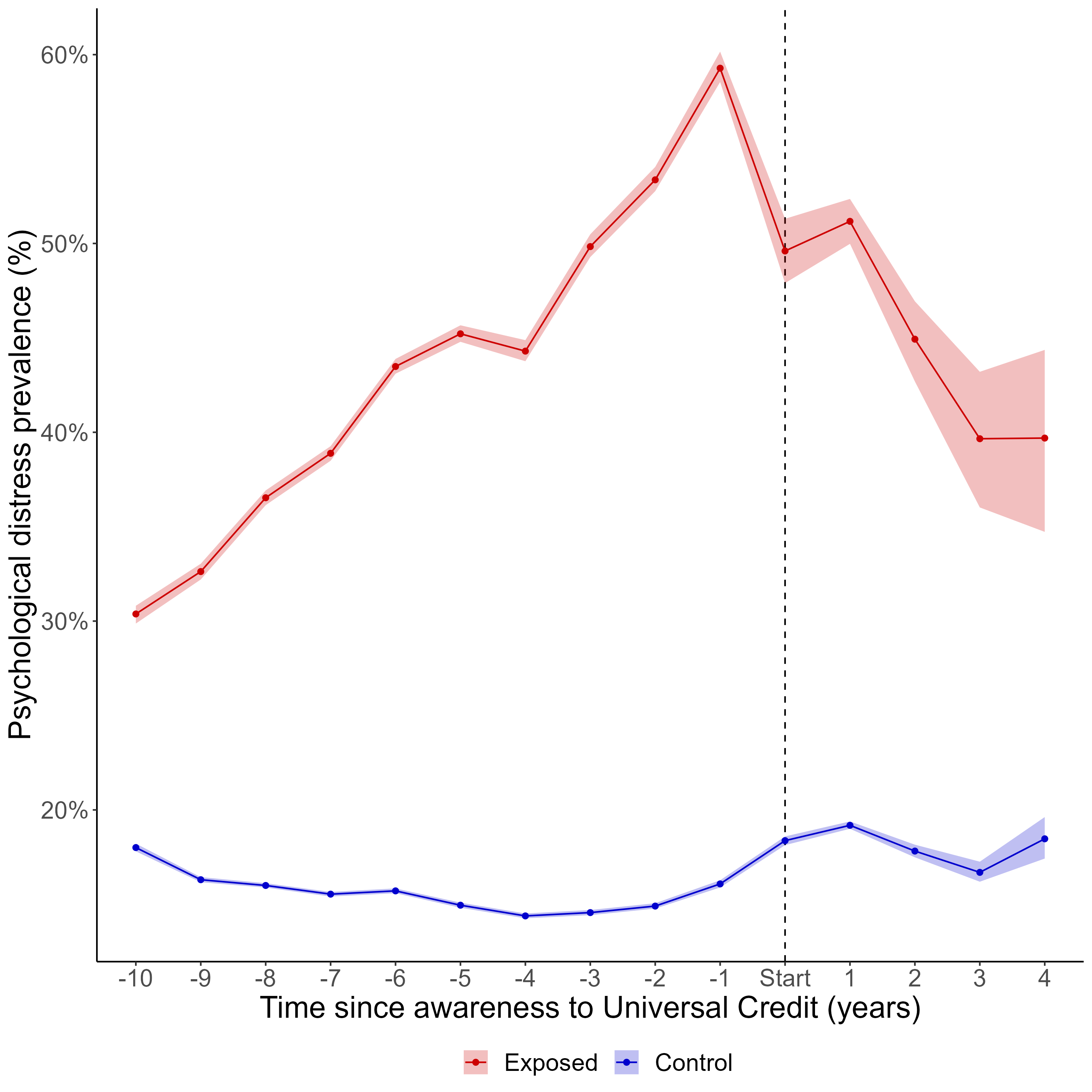}
\end{figure}

Figure \ref{Fig: centeredPlot.national} shows the national prevalence of self-reported psychological distress by exposure group over the study period. The prevalence of psychological distress in the exposed population is both higher and more variable than the prevalence of the control population. The exposed population sees an increasing trend in the prevalence of psychological distress during the years prior to the intervention, culminating in a global peak in the year immediately prior to the intervention. From this point, the prevalence decreases slightly before plateauing at a higher level than when it started. In comparison, the control population remains relatively stable and sees a small increase in the year immediately following the intervention. Figure \ref{Fig: centeredPlot.national} not only highlights the disparity in the prevalence between the two populations but highlights how the build up to the intervention has vastly difference effects on them.

\begin{table*}[!ht]
    \centering
    \caption{Psychological distress prevalence for the exposed and control populations, and the difference between them, for different temporal profiles.}
    \label{Tab: Temporal Profile Summary Results}
    \resizebox{\textwidth}{!}{%
    \begin{tabular}{c|ccc|ccc|ccc}
        \hline
        \multirow{2}{*}{\textbf{Temporal level}} & \multicolumn{3}{c|}{\textbf{Exposed prevalence (\%)}} & \multicolumn{3}{c|}{\textbf{Control prevalence (\%)}} & \multicolumn{3}{c}{\textbf{Exposed / Control}} \\ 
        & \textbf{Lower} & \textbf{Median} & \textbf{Upper} & \textbf{Lower} & \textbf{Median} & \textbf{Upper} & \textbf{Lower} & \textbf{Median} & \textbf{Upper} \\ \hline
        $\texttt{Year} = -10$ & 29.87 & 30.38 & 30.81 & 17.81 & 18.01 & 18.22 & 1.66 & 1.69 & 1.72 \\
        $\texttt{Year} = -9$ & 32.20 & 32.62 & 33.04 & 16.17 & 16.30 & 16.44 & 1.97 & 2.00 & 2.03 \\
        $\texttt{Year} = -8$ & 36.14 & 36.53 & 36.92 & 15.89 & 16.00 & 16.12 & 2.26 & 2.28 & 2.31 \\
        $\texttt{Year} = -7$ & 38.49 & 38.89 & 39.26 & 15.41 & 15.54 & 15.66 & 2.48 & 2.50 & 2.53 \\
        $\texttt{Year} = -6$ & 43.10 & 43.48 & 43.89 & 15.59 & 15.71 & 15.84 & 2.74 & 2.77 & 2.79 \\
        $\texttt{Year} = -5$ & 44.79 & 45.22 & 45.67 & 14.82 & 14.95 & 15.09 & 3.00 & 3.02 & 3.06 \\
        $\texttt{Year} = -4$ & 43.76 & 44.30 & 44.88 & 14.26 & 14.39 & 14.51 & 3.04 & 3.08 & 3.12 \\
        $\texttt{Year} = -3$ & 49.28 & 49.83 & 50.49 & 14.43 & 14.56 & 14.70 & 3.38 & 3.42 & 3.47 \\
        $\texttt{Year} = -2$ & 52.77 & 53.37 & 54.05 & 14.77 & 14.91 & 15.06 & 3.53 & 3.58 & 3.64 \\
        $\texttt{Year} = -1$ & 58.55 & 59.28 & 60.16 & 15.89 & 16.08 & 16.28 & 3.63 & 3.69 & 3.74 \\
        $\texttt{Year} = 0$ & 47.90 & 49.60 & 51.31 & 18.14 & 18.37 & 18.60 & 2.60 & 2.70 & 2.80 \\
        $\texttt{Year} = 1$ & 49.98 & 51.17 & 52.36 & 19.00 & 19.18 & 19.39 & 2.60 & 2.67 & 2.73 \\
        $\texttt{Year} = 2$ & 42.69 & 44.93 & 46.94 & 17.49 & 17.82 & 18.16 & 2.39 & 2.52 & 2.64 \\
        $\texttt{Year} = 3$ & 36.02 & 39.66 & 43.20 & 16.20 & 16.69 & 17.26 & 2.15 & 2.37 & 2.58 \\
        $\texttt{Year} = 4$ & 34.73 & 39.69 & 44.36 & 17.42 & 18.47 & 19.62 & 1.86 & 2.14 & 2.42 \\
        Before intervention & 41.59 & 41.90 & 42.25 & 15.35 & 15.42 & 15.50 & 2.69 & 2.72 & 2.74 \\
        After intervention & 48.38 & 49.51 & 50.68 & 18.36 & 18.53 & 18.72 & 2.61 & 2.67 & 2.73 \\
        All years & 42.39 & 42.73 & 43.07 & 15.86 & 15.93 & 16.01 & 2.66 & 2.68 & 2.70 \\ \hline
    \end{tabular}%
    }
\end{table*}

To give a quantitative assessment of the prevalence of psychological distress, Table \ref{Tab: Temporal Profile Summary Results} shows a summary of results for the control and exposed populations at different temporal profiles. The top rows are temporal profiles defined by $\texttt{Year}$ and the bottom three rows are defined by aggregating the posterior distribution over different combinations of years to form all the years before and after the intervention and all the years in the study period. The first six columns are the median and 95\% CrI for the exposed and control population, respectively, and the final three columns are the median and 95\% CrI for the posterior distribution defined by the ratio between the exposed and control populations at each temporal profile. This ratio is not the standardised change ($\rho$; which considers the difference in the exposed population due to the intervention once adjusted for the change in the control population due to the intervention) but is used as a measure of the disparity between the two populations at the different temporal profiles. 

Across the study period, the disparity (i.e., ratio) between the two populations is steadily increasing in the years building up to the intervention and peaks in the year prior, 3.69 (95\% CrI: 3.63--3.74). Whilst the disparity between the two populations becomes slightly smaller in the years after the intervention when compared to the years before, the overall disparity across the whole study period is that the prevalence in the exposed population is on average 2.68 (95\% CrI: 2.66--2.71) higher than in the control population. 

\subsection{Standardised change by spatial profile}

\begin{figure}[!h]
    \centering
    \caption{Standardised change due to a contextual awareness of Universal Credit for each Lower Tier Local Authority (LTLA): (a) map showing the distribution by geographical location; (b) 95\% Credible Intervals ordered by increases in psychological distress, including the national change.}
    \label{Fig: Spatial Standardised Change Results}    
    \begin{subfigure}[t]{0.48\textwidth}
        \centering
        \caption{Median standardised change}
        \label{Fig: standChangePlot.localAuthority_prev}
        \includegraphics[width=0.85\textwidth]{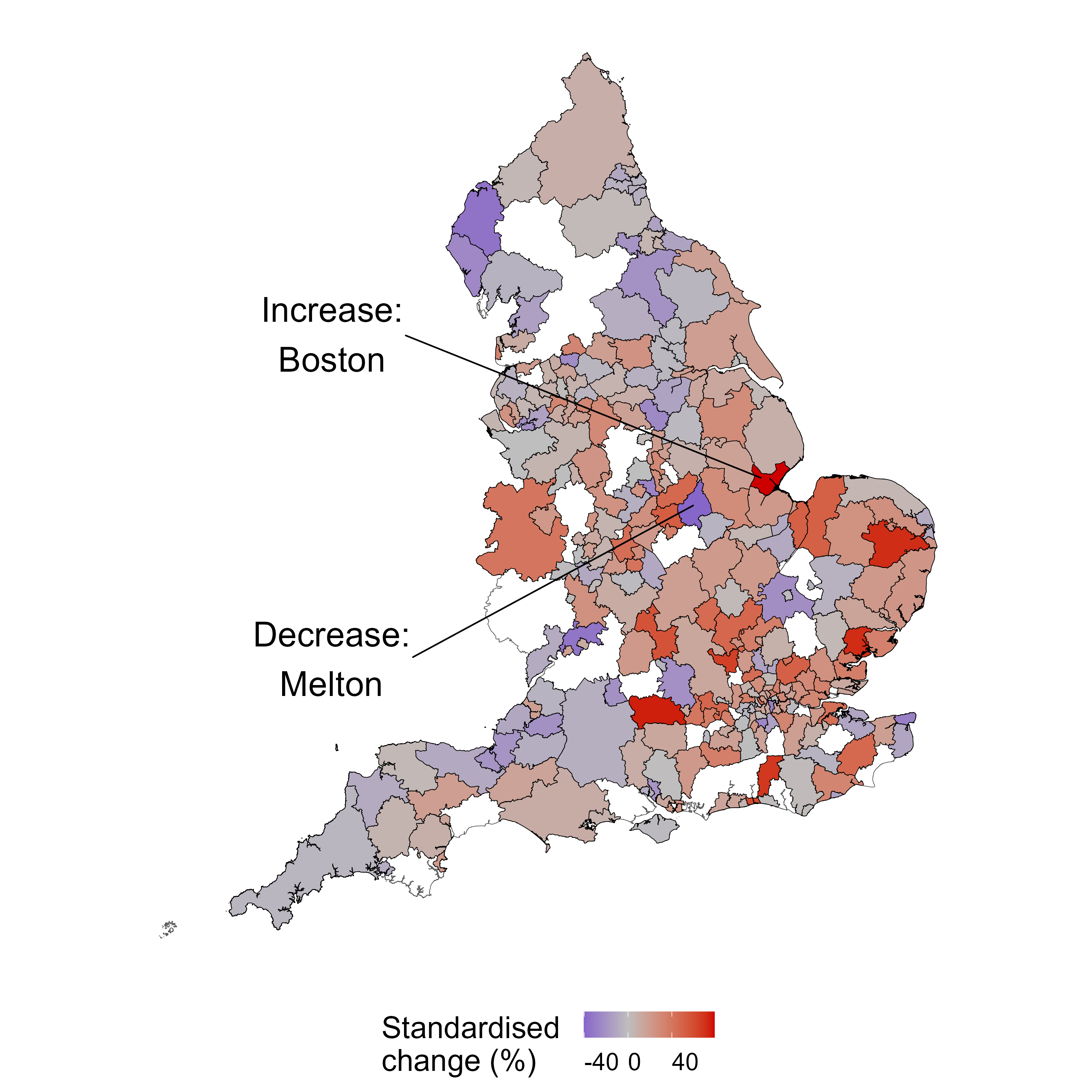}
    \end{subfigure}%
    \begin{subfigure}[t]{0.48\textwidth}
        \centering
        \caption{Uncertainty of standardised change}
        \label{Fig: standChangePlot.localAuthority_uncer}
        \includegraphics[width=0.85\textwidth]{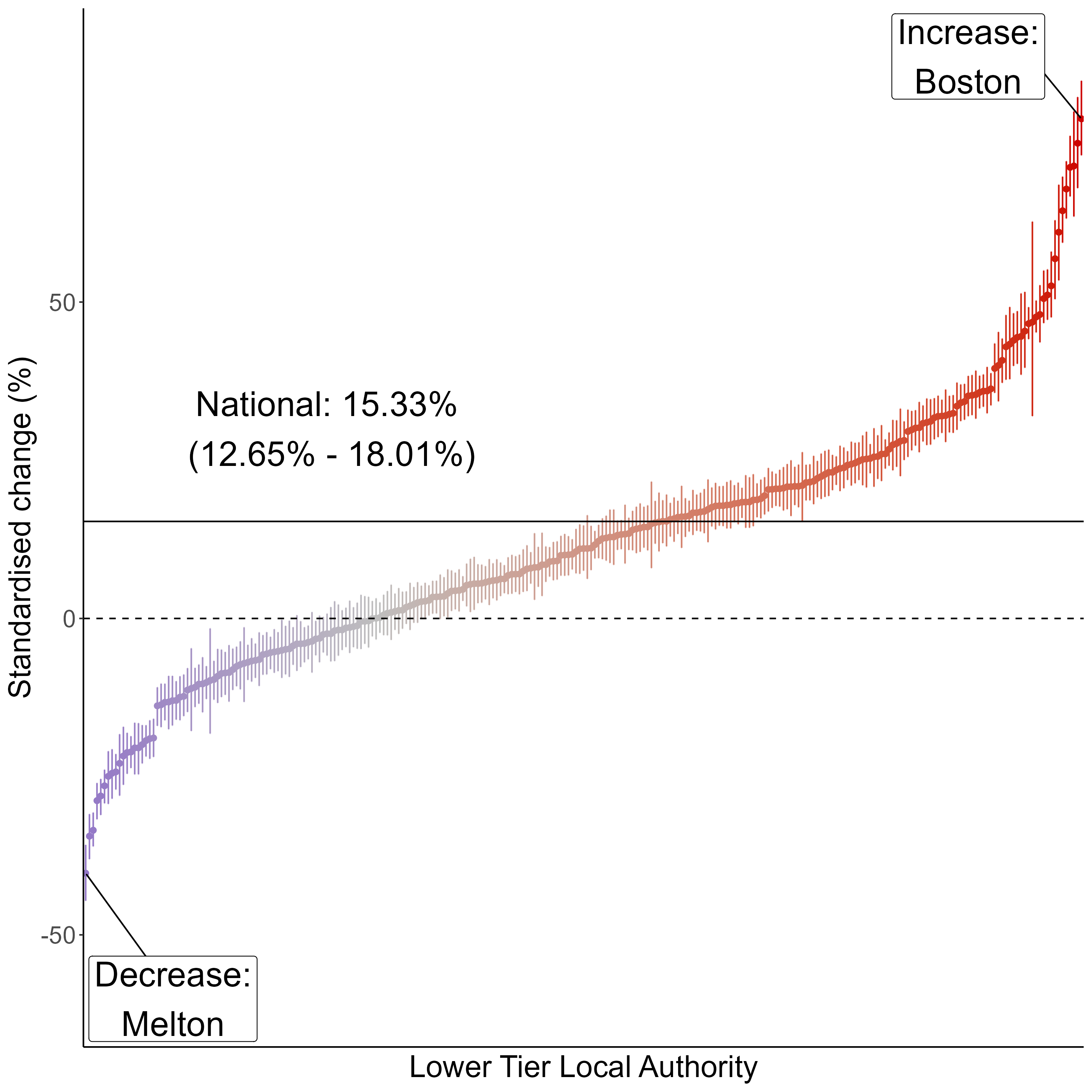}
    \end{subfigure}    
\end{figure}

Figure \ref{Fig: Spatial Standardised Change Results} shows the standardised change of psychological distress in the exposed population due to a contextual awareness of UC for each LTLA. The standardised change was calculated using all years in the study period (2009--2021). Note that in the map depicted in Fig (\ref{Fig: standChangePlot.localAuthority_prev}), there are 44 missing LTLAs (shown in white). This is as they do not have participants in the UKHLS belonging to either the exposed or control group, before or after. Most of the missing data comes from LTLAs missing an exposed individual after the intervention (42 LTLAs are missing this). 

The large, uncorrelated differences between LTLAs in the spatial map provides evidence that the choice of an unstructured spatial effect is suitable since there is residual, unstructured spatial variation being captured by the model. In addition, the uncertainty plot \ref{Fig: standChangePlot.localAuthority_uncer} shows the range of change for all LTLAs, highlighting the LTLAs with the largest increase and decrease in psychological distress.

In addition to the change for each LTLA, Figure \ref{Fig: standChangePlot.localAuthority_uncer} includes the national standardised change, 15.30\% (95\% CrI: 12.60\%--18.30\%), which provides substantial evidence that a contextual awareness of UC had a negative impact on mental well-being on those in England who are unemployed. Our results that UC had a negative impact on those unemployed in England is in-line with previous results \citep{wickham2020effects}. However, we extended the results by including additional insights at the subnational level where is clear, the overall increase seen at the national level is not reflected in all LTLAs. 

\subsection{Standardised change by socioeconomic profiles}

\begin{figure}[!h]
    \centering
    \caption{Plot of the standardised changes for each category of the individual and community confounder profiles: median and error bars (95\% CrI). The dashed black line indicates 0\% standardised change.}
    \label{Fig: standChangePlot.allConfounders}
    \includegraphics[width=\textwidth]{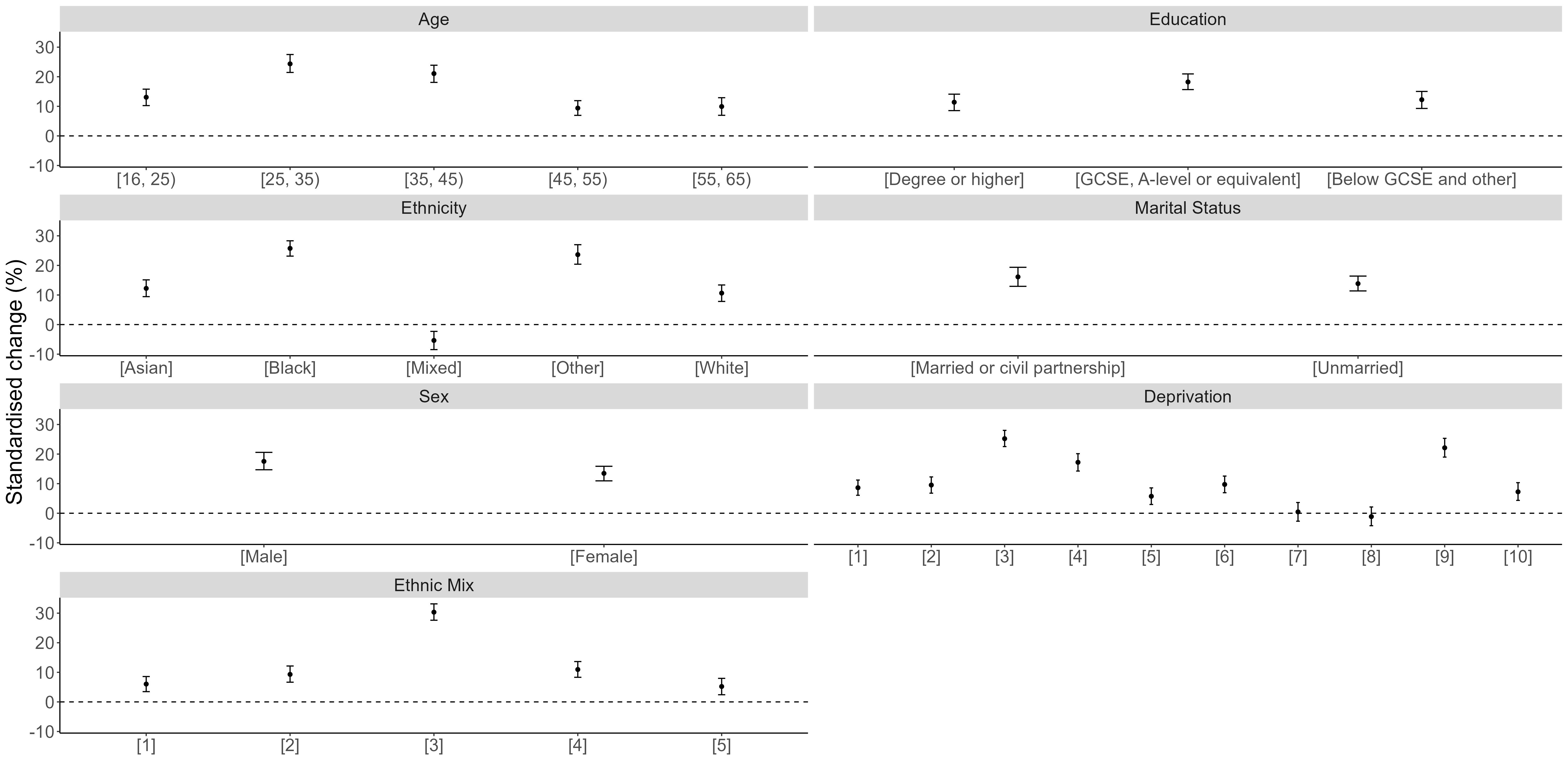}
\end{figure}

Figure \ref{Fig: standChangePlot.allConfounders} shows the standardised change for each confounder category individually. For all but the [Mixed] ethnic group and deprivation deciles [7] and [8], the intervention caused an increase in the prevalence of psychological distress.

\begin{figure}[!h]
    \centering
    \caption{Tile maps of joint standardised change for individual and community profiles. The left- and right-hand columns are for the individual confounders joint with the deprivation ([1] being the most deprived and [10] the least) and ethnic mix ([1] being the largest ethnic mix and [5] the least) confounders, respectively. The rows represent individual confounders (age, education, ethnicity, marital status and sex from top-to-bottom). The blue-red colour change represents a decrease-increase in the standardised change for before-after the intervention.}
    \label{Fig: standChangePlot.jointConfounders}
    \includegraphics[width=0.5\textwidth]{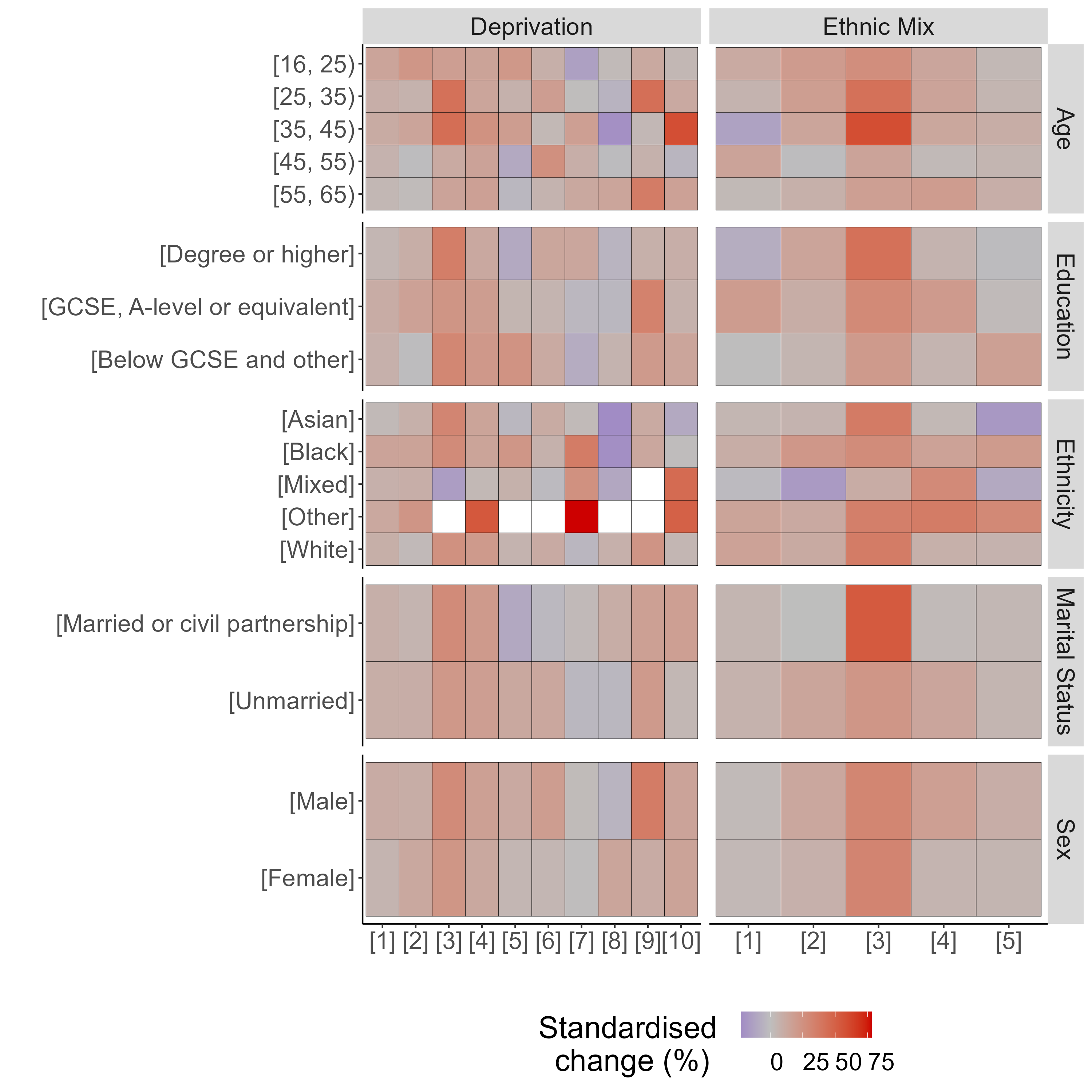}
\end{figure}

In addition to the standardised change for each confounder category individually, we produced the standardised change for joint individual (age, education, ethnicity, marital status, and sex) and community (deprivation and ethnic mix) confounder categories. The results are presented in Figure \ref{Fig: standChangePlot.jointConfounders}. The plot shows most combinations of individual and community confounders have either no change (grey) or an increase in psychological distress (red) in the years following the intervention. There are a few cases of a decrease in psychological distress (lighter blue), and these are generally found in less deprived areas (columns [7] to [10] on the x-axis in Figure \ref{Fig: standChangePlot.jointConfounders}).

Table \ref{Tab: Top and Bottom 5 Profiles} shows the combinations of individual level confounders with the top five largest increase and decrease in psychological distress due to the intervention. For example, the largest increase is estimated for individuals with the following characteristics: $\rOpenInt{16, 25}$, [GCSE, A-level or equivalent], [Black], and [Unmarried]. The results in Table \ref{Tab: Top and Bottom 5 Profiles} can be seen as a combination of the results in Figure \ref{Fig: standChangePlot.allConfounders}. For example, the younger ages, [Male] and [Unmarried] featured more heavily in the top five increases in comparison to top five decreases. Furthermore, those categories had the largest marginalised standardised change in each of the confounders in Figure  \ref{Fig: standChangePlot.allConfounders}. 

The results shown in this section are used to highlight the range and wealth of results available from the \bhf \ framework. The rich model output allows to define marginal posteriors distributions for each variable or random effect separately, as well as profiles of individual level and community levels characteristics. Additionally, summaries can be obtained from the posterior distribution together with posterior distribution of functions of the parameters (e.g. $\rho$) and the corresponding uncertainty. 

\begin{table}[!h]
    \centering
    \caption{Five profile combinations with the largest increase (top five rows) and decrease (bottom five rows) in standardised change.}
    \label{Tab: Top and Bottom 5 Profiles}
    \resizebox{\textwidth}{!}{%
        \begin{tabular}{c|c|c|c|c|ccc}
            \hline
            \multirow{2}{*}{\textbf{Age}} & \multirow{2}{*}{\textbf{Education}} & \multirow{2}{*}{\textbf{Ethnicity}} & \multirow{2}{*}{\textbf{Marital Status}} & \multirow{2}{*}{\textbf{Sex}} & \multicolumn{3}{c}{\textbf{Standardised change (\%)}} \\
            &  &  &  &  & \textbf{Lower} & \textbf{Median} & \textbf{Upper} \\ \hline
            $\rOpenInt{16, 25}$ & [GCSE, A-level or equivalent] & [Black] & [Unmarried] & [Male] & 51.01 & 55.96 & 60.69 \\
            $\rOpenInt{25, 35}$ & [Degree or higher] & [White] & [Married or civil partnership] & [Female] & 42.13 & 51.22 & 60.13 \\
            $\rOpenInt{25, 35}$ & [Below GCSE and other] & [Black] & [Unmarried] & [Female] & 42.67 & 47.00 & 50.73 \\
            $\rOpenInt{45, 55}$ & [GCSE, A-level or equivalent] & [Black] & [Married or civil partnership] & [Male] & 39.95 & 46.25 & 52.25 \\
            $\rOpenInt{25, 35}$ & [Below GCSE and other] & [Other] & [Unmarried] & [Male] & 38.14 & 45.51 & 54.62 \\
            \dots & \dots & \dots & \dots & \dots & \dots & \dots & \dots \\ 
            $\rOpenInt{55, 65}$ & [Degree or higher] & [Mixed] & [Married or civil partnership] & [Male] & -31.82 & -22.19 & -13.34 \\
            $\rOpenInt{45, 55}$ & [GCSE, A-level or equivalent] & [Asian] & [Married or civil partnership] & [Male] & -26.15 & -22.91 & -19.73 \\
            $\rOpenInt{55, 65}$ & [Below GCSE and other] & [Black] & [Unmarried] & [Male] & -27.97 & -23.78 & -20.05 \\
            $\rOpenInt{45, 55}$ & [Degree or higher] & [White] & [Married or civil partnership] & [Female] & -26.70 & -24.29 & -22.09 \\
            $\rOpenInt{35, 45}$ & [Below GCSE and other] & [Black] & [Married or civil partnership] & [Male] & -32.46 & -25.45 & -17.21 \\ \hline
        \end{tabular}%
    }
\end{table}

\subsection{Sensitivity analysis}

\begin{figure}[!h]
    \centering
    \caption{Sensitivity analysis for the definition of when the intervention occurs. The left plot is for the exposed population and the right plot is for the control population. The line in red is when the intervention is defined to have started when an LTLA is introduced to Universal Credit. The green, orange, blue, purple, and pink lines are when the intervention is defined by a contextual awareness where the threshold is set as 5\%, 15\%, 25\%, 35\% and 45\%, respectively, of the most recent number of people on Universal Credit.}
    \label{Fig: sensitivityAnalysis1}
    \includegraphics[width=\textwidth]{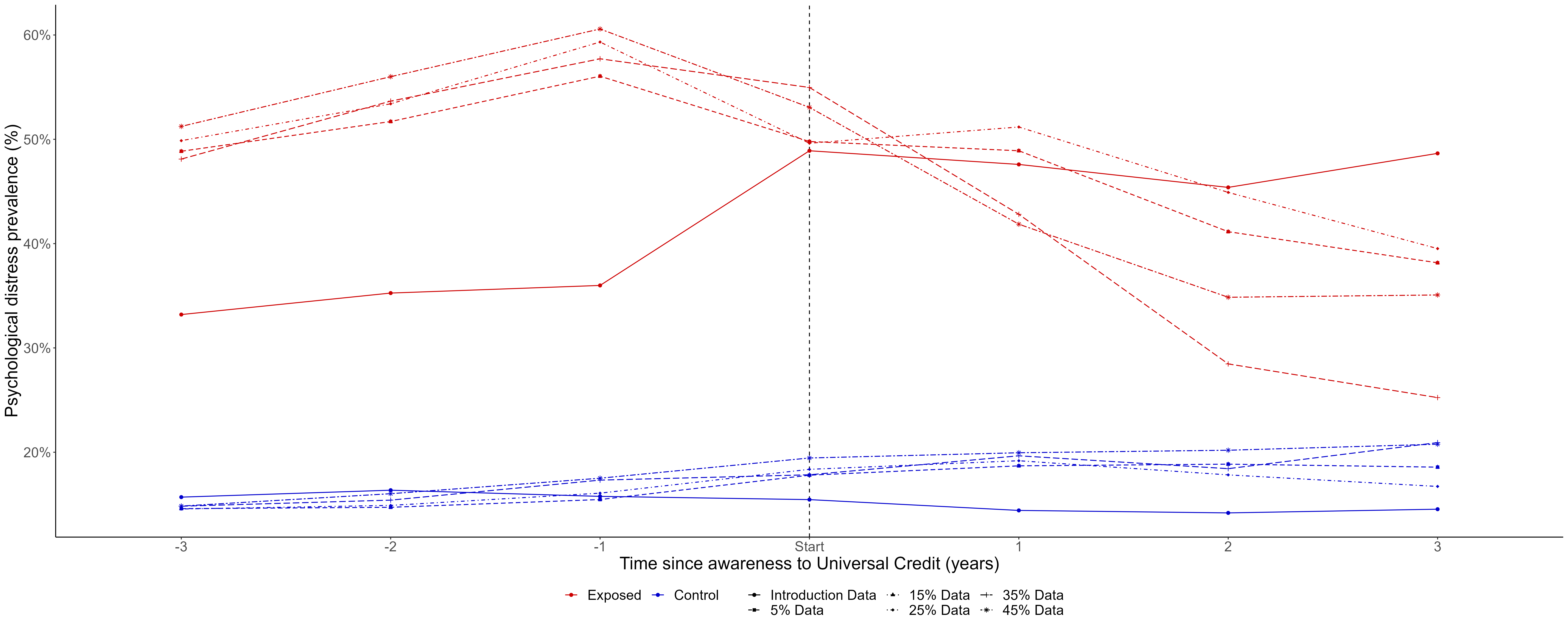}
\end{figure}

We ran a sensitivity analysis on the definition of a contextual awareness to Universal Credit (UC); the results are shown in Figure \ref{Fig: sensitivityAnalysis1}. We compared the national average of self-reported psychological distress for the exposed and control populations (i) when the intervention is defined when at least one person in an LTLA is receiving UC (introduction), (ii) when the threshold for contextual awareness is set to 5\%, 15\%, 25\%, 35\% and 45\% of the most recent number of people on UC. 

In Figure \ref{Fig: sensitivityAnalysis1}, there are differences between each data set implying the choice in definition of when the intervention has begun is important. There is a clear difference in the pattern within the exposed and control population between the sets of data where the intervention was defined by the contextual awareness and where it was defined by an introduction. For the exposed population, the contextual awareness data peaks in the year before the intervention for all (5\%, 15\%, 25\%, 35\% and 45\%) and then declines, whereas the introduction data jumps up the year of the intervention and remains relatively flat. In the control population, the contextual awareness data increases slightly from the year prior to the intervention and then remains flat whereas the introduction data has a slight decrease the year the intervention occurs. In both the exposed and control populations, the patterns in the results for the data defined by a contextual awareness are more similar to one another than the pattern in the introduction data. However, the main take away from the sensitivity analysis is the definition of when the intervention starts can have a substantial impact on the results and their interpretation.

\section{Discussion}

In this article, we presented a flexible and generalisable ITS model in a Bayesian framework, which can be employed to evaluate policy interventions, incorporating variables at different scales, and including spatial and temporal random effects if required, to fully account for heterogeneity. We investigated differences at the subnational level, which is crucial when evaluating policy, as geography is likely to be a proxy for residual confounding in studies of this kind. A unique advantage of the proposed framework is that profiles identified by set of covariates can be queried to assess and compare result, flagging potential inequalities.

To demonstrate our framework, we explored the effect of a contextual awareness to UC on self-reported psychological distress for exposed and control groups based on self-reported employment response. Overall, in England, UC caused a 15.30\% (95\% CrI: 12.60\%--18.30\%) increase in psychological distress in the exposed population (after adjustment for the change in the control population). We showed the ratio between the prevalence of psychological distress in the exposed and control populations over the entire study period was 2.68 (95\% CrI: 2.66--2.71), with a peak of 3.69 (95\% CrI: 3.63--3.74) in the year immediately prior to the intervention. The peak reflects the definition of contextual awareness where individuals’ mental well-being changes in anticipation of the intervention. When considering the effect at a subnational level, we highlighted the large variation between different LTLAs with some having an increase in psychological distress within its exposed population and others having a decrease. 

Our approach builds upon the work of \citet{wickham2020effects} who use a difference-in-difference (DID) design that adjusts for individual level confounders (not including ethnicity) implemented using a frequentist paradigm to assess the impact of an introduction to UC on mental well-being for exposed and control groups based on self-reported employment responses. \citet{wickham2020effects} found that due to UC, the prevalence of psychological distress in the intervention group increased by 21\% relative to its baseline. The analogue in our analysis would be the national standardised change in psychological distress for the unemployed, 15.30\% (95\% CrI: 12.60\%--18.30\%). Differences can be attributed (i) to methodological aspects (e.g., the type of quasi-experimental design), the covariates included in the model and the weights and, (ii) difference in data sets (they used periods 2009--2018 and we used periods 2009--2021). 

A DID only considers one measurement before and after an intervention and does not include a temporal trend; consequently, we used an ITS as it is better suited for interventions in time \citep{wagner2002segmented, penfold2013use}. In addition, we included additional extensions to account for residual confounding in space and confounding due to community level deprivation and ethnicity. Using the strengths of our methodology, we produced results that provide a deeper insight into the effect of UC on mental well-being than is currently available in the literature. For example, the large variation in the change of psychological distress in the exposed population for different LTLAs is masked when considering the national results only. Furthermore, the exploration of different profiles is important to understand the influence of UC on certain characteristics to identify those most at risk. Insights into differences between the profiles is invaluable for effective data-driven policy evaluation and implementation. For example, it highlights combinations of location, community- and individual-level characteristics that would benefit from additional support during similar interventions thereby moving towards a more effective and efficient implementation. 

During 2017--2022, of the people on UC approximately 40\% of them are in employment \citepalias{dwp2022ucStatistics}. Therefore, due to our choice to use employment as a proxy for exposure to UC (there was insufficient individual level data on UC from the UKHLS), the results for the exposed and control group will be under- and over-estimated, respectively. The lack of robustness in the results of the sensitivity analysis highlighted how the definition of the start of the intervention is an extremely important choice. For future researchers who are considering other policies at the population level where there is not information available at the individual level, we recommend caution and to perform a sensitivity analysis to understand any implications. An extension we could make is to consider additional outcomes. For example, we used a dichotomous outcome derived from the GHQ-12 scores. However, we could use the GHQ-12 scores themselves, or other continuous measures of psychological distress from the UKHLS, i.e., the mental component summary that gives a continuous outcome of active depression based on a 12-item Short Form Health Survey.

To conclude, we believe that the Bayesian hierarchical framework is the natural approach for evaluating the impact of policy interventions at population level, taking advantage of the intrinsic longitudinal nature of the data as well as of spatial and temporal dependencies. The framework we provided is not solely applicable for changes in mental well-being due to UC; for example, the same framework could be used assess the impact of the other policy changes in the UK on mental health outcomes, i.e., the impact of the UK's 2012 Suicide Prevention Strategy on suicide rates. Alternatively, it can be used more generally on ``shocks'' in time that affect health outcomes, i.e., the impact of COVID-19.

\section*{Data Statement}

From the UKHLS, the main survey is free to download after registration, but for data sets that contain geographical information, special licences are required \citepalias{ukhls2022data}. Data on when a LTLA was made contextually aware to UC is free to download from the DWP after registration \citepalias{dwp2022statxplore}. 

Data on the IMD \citepalias{ons2019imd}, proportion of the population from a Black, Asian or other minority ethnic groups \citepalias{ons2022bame}, and the spatial shapefiles of England \citepalias{ons2023geog} are free to access and download from public resources.

The full code for implementing the analysis in this paper can be found at \url{https://github.com/connorgascoigne/Bayesian-ITS-for-policy}.

\clearpage

\bibliographystyle{abbrvnat}
\bibliography{library}

\begin{thebibliography}{40}
\providecommand{\natexlab}[1]{#1}
\providecommand{\url}[1]{\texttt{#1}}
\expandafter\ifx\csname urlstyle\endcsname\relax
  \providecommand{\doi}[1]{doi: #1}\else
  \providecommand{\doi}{doi: \begingroup \urlstyle{rm}\Url}\fi

\bibitem[Abadie and Gardeazabal(2003)]{abadie2003economic}
A.~Abadie and J.~Gardeazabal.
\newblock The economic costs of conflict: A case study of the basque country.
\newblock \emph{American economic review}, 93\penalty0 (1):\penalty0 113--132,
  2003.

\bibitem[Barnes and Bates(2017)]{barnes2017racial}
D.~M. Barnes and L.~M. Bates.
\newblock Do racial patterns in psychological distress shed light on the
  black--white depression paradox? a systematic review.
\newblock \emph{Social psychiatry and psychiatric epidemiology}, 52:\penalty0
  913--928, 2017.

\bibitem[Bernal et~al.(2017)Bernal, Cummins, and
  Gasparrini]{bernal2017interrupted}
J.~L. Bernal, S.~Cummins, and A.~Gasparrini.
\newblock Interrupted time series regression for the evaluation of public
  health interventions: a tutorial.
\newblock \emph{International journal of epidemiology}, 46\penalty0
  (1):\penalty0 348--355, 2017.

\bibitem[Byrne et~al.(2020)Byrne, Alexander, Khan, Nazroo, and
  Shankley]{byrne2020ethnicity}
B.~Byrne, C.~Alexander, O.~Khan, J.~Nazroo, and W.~Shankley.
\newblock \emph{Ethnicity, Race and Inequality in the UK: State of the Nation}.
\newblock Policy press, 2020.

\bibitem[Campbell and Stanley(2015)]{campbell2015experimental}
D.~T. Campbell and J.~C. Stanley.
\newblock \emph{Experimental and quasi-experimental designs for research}.
\newblock Ravenio books, 2015.

\bibitem[Cheetham et~al.(2022)Cheetham, Atkinson, Gibson, Katikireddi, Moffatt,
  Morris, Munford, Shenton, Wickham, and Craig]{cheetham2022exploring}
M.~Cheetham, P.~Atkinson, M.~Gibson, S.~Katikireddi, S.~Moffatt, S.~Morris,
  L.~Munford, F.~Shenton, S.~Wickham, and P.~Craig.
\newblock Exploring the mental health effects of universal credit: a journey of
  co-production.
\newblock \emph{Perspectives in public health}, 142\penalty0 (4):\penalty0
  209--212, 2022.

\bibitem[Cook et~al.(2002)Cook, Campbell, and Shadish]{cook2002experimental}
T.~D. Cook, D.~T. Campbell, and W.~Shadish.
\newblock \emph{Experimental and quasi-experimental designs for generalized
  causal inference}.
\newblock Houghton Mifflin Boston, MA, 2002.

\bibitem[Corris et~al.(2020)Corris, Dormer, Brown, Whitty, Collingwood, Bambra,
  and Newton]{corris2020health}
V.~Corris, E.~Dormer, A.~Brown, P.~Whitty, P.~Collingwood, C.~Bambra, and J.~L.
  Newton.
\newblock Health inequalities are worsening in the north east of england.
\newblock \emph{British Medical Bulletin}, 134\penalty0 (1):\penalty0 63--72,
  2020.

\bibitem[Craig and Katikireddi(2020)]{craig2020early}
P.~Craig and S.~V. Katikireddi.
\newblock Early impacts of universal credit: the tip of the iceberg?
\newblock \emph{The Lancet Public Health}, 5\penalty0 (3):\penalty0 e131--e132,
  2020.

\bibitem[{Department for Levelling Up, Housing and
  Communities}(2016)]{gov2016ltla}
{Department for Levelling Up, Housing and Communities}.
\newblock \emph{Local government structure and elections}.
\newblock {Department for Work and Pensions},
  \url{https://www.gov.uk/guidance/local-government-structure-and-elections},
  2016.
\newblock {Accessed May 2023}.

\bibitem[{Department for Work and Pensions}(2022 (a))]{dwp2022statxplore}
{Department for Work and Pensions}.
\newblock \emph{{Stat-Xplore}}.
\newblock \url{https://stat-xplore.dwp.gov.uk}, 2022 (a).
\newblock {Accessed May 2023}.

\bibitem[{Department for Work and Pensions}(2022 (b))]{dwp2022ucStatistics}
{Department for Work and Pensions}.
\newblock \emph{Universal Credit statistics, 29 April 2013 to 14 July 2022}.
\newblock
  \url{https://www.gov.uk/government/statistics/universal-credit-statistics-29-april-2013-to-14-july-2022},
  2022 (b).
\newblock {Accessed May 2023}.

\bibitem[{Department for Work and Pensions}(2022 (c))]{dwp2022ucAge}
{Department for Work and Pensions}.
\newblock \emph{Universal Credit}.
\newblock \url{https://www.gov.uk/universal-credit/eligibility}, 2022 (c).
\newblock {Accessed December 2022}.

\bibitem[Dotsikas et~al.(2023)Dotsikas, Osborn, Walters, and
  Dykxhoorn]{dotsikas2023trajectories}
K.~Dotsikas, D.~Osborn, K.~Walters, and J.~Dykxhoorn.
\newblock Trajectories of housing affordability and mental health problems: a
  population-based cohort study.
\newblock \emph{Social Psychiatry and Psychiatric Epidemiology}, 58\penalty0
  (5):\penalty0 769--778, 2023.

\bibitem[Ellis and Fry(2010)]{ellis2010regional}
A.~Ellis and R.~Fry.
\newblock Regional health inequalities in england.
\newblock \emph{Regional Trends}, 42:\penalty0 60--79, 2010.

\bibitem[Forth(2021)]{forth2021regional}
T.~Forth.
\newblock Regional inequalities post-brexit: Levelling-up.
\newblock \emph{UK in a Changing Europe}, 2021.

\bibitem[Freni-Sterrantino et~al.(2019)Freni-Sterrantino, Ghosh, Fecht,
  Toledano, Elliott, Hansell, and Blangiardo]{freni2019bayesian}
A.~Freni-Sterrantino, R.~Ghosh, D.~Fecht, M.~Toledano, P.~Elliott, A.~Hansell,
  and M.~Blangiardo.
\newblock Bayesian spatial modelling for quasi-experimental designs: An
  interrupted time series study of the opening of municipal waste incinerators
  in relation to infant mortality and sex ratio.
\newblock \emph{Environment international}, 128:\penalty0 109--115, 2019.

\bibitem[Gnambs and Staufenbiel(2018)]{gnambs2018structure}
T.~Gnambs and T.~Staufenbiel.
\newblock The structure of the general health questionnaire (ghq-12): two
  meta-analytic factor analyses.
\newblock \emph{Health psychology review}, 12\penalty0 (2):\penalty0 179--194,
  2018.

\bibitem[{Health Survey England}(2017)]{hes2017well}
{Health Survey England}.
\newblock \emph{Well-being and mental health, 2016}.
\newblock
  \url{http://healthsurvey.hscic.gov.uk/support-guidance/public-health/health-survey-for-england-2016/well-being-and-mental-health.aspx},
  2017.
\newblock {Accessed May 2023}.

\bibitem[{Institute of Fiscal Studies}(2020)]{ifs2020catching}
{Institute of Fiscal Studies}.
\newblock \emph{Catching up or falling behind? Geographical inequalities in the
  UK and how they have changed in recent years}.
\newblock
  \url{https://ifs.org.uk/inequality/geographical-inequalities-in-the-uk/},
  2020.
\newblock {Accessed May 2023}.

\bibitem[Linden and Adams(2011)]{linden2011applying}
A.~Linden and J.~L. Adams.
\newblock Applying a propensity score-based weighting model to interrupted time
  series data: improving causal inference in programme evaluation.
\newblock \emph{Journal of evaluation in clinical practice}, 17\penalty0
  (6):\penalty0 1231--1238, 2011.

\bibitem[Lorant et~al.(2018)Lorant, De~Gelder, Kapadia, Borrell, Kalediene,
  Kov{\'a}cs, Leinsalu, Martikainen, Menvielle, Regidor,
  et~al.]{lorant2018socioeconomic}
V.~Lorant, R.~De~Gelder, D.~Kapadia, C.~Borrell, R.~Kalediene, K.~Kov{\'a}cs,
  M.~Leinsalu, P.~Martikainen, G.~Menvielle, E.~Regidor, et~al.
\newblock Socioeconomic inequalities in suicide in europe: the widening gap.
\newblock \emph{The British Journal of Psychiatry}, 212\penalty0 (6):\penalty0
  356--361, 2018.

\bibitem[Mahase(2020)]{mahase2020universal}
E.~Mahase.
\newblock \emph{Universal credit linked to psychological distress but not
  employment}, 2020.

\bibitem[Marmot(2020)]{marmot2020health}
M.~Marmot.
\newblock {Health equity in England: the Marmot review 10 years on}.
\newblock \emph{British Medical Journal}, 368\penalty0 (1), 2020.

\bibitem[Muntaner et~al.(2004)Muntaner, Eaton, Miech, and
  O’campo]{muntaner2004socioeconomic}
C.~Muntaner, W.~W. Eaton, R.~Miech, and P.~O’campo.
\newblock {Socioeconomic position and major mental disorders}.
\newblock \emph{Epidemiologic reviews}, 26\penalty0 (1):\penalty0 53--62, 2004.

\bibitem[{Office for National Statistics}(2019)]{ons2019imd}
{Office for National Statistics}.
\newblock \emph{English indices of deprivation 2019}.
\newblock
  \url{https://www.gov.uk/government/statistics/english-indices-of-deprivation-2019},
  2019.
\newblock {Accessed December 2022}.

\bibitem[{Office for National Statistics}(2020)]{ons2021census}
{Office for National Statistics}.
\newblock \emph{Census 2021 geographies}.
\newblock
  \url{https://www.ons.gov.uk/methodology/geography/ukgeographies/censusgeographies/census2021geographies},
  2020.
\newblock {Accessed May 2023}.

\bibitem[{Office for National Statistics}(2022 (a))]{ons2022bame}
{Office for National Statistics}.
\newblock \emph{Population of England and Wales}.
\newblock
  \url{https://www.ethnicity-facts-figures.service.gov.uk/uk-population-by-ethnicity/national-and-regional-populations/population-of-england-and-wales/latest},
  2022 (a).
\newblock {Accessed December 2022}.

\bibitem[{Office for National Statistics}(2023)]{ons2023geog}
{Office for National Statistics}.
\newblock \emph{The Open Geography portal}.
\newblock \url{https://geoportal.statistics.gov.uk/}, 2023.
\newblock {Accessed May 2023}.

\bibitem[O’Donoghue et~al.(2016)O’Donoghue, Roche, and
  Lane]{o2016neighbourhood}
B.~O’Donoghue, E.~Roche, and A.~Lane.
\newblock {Neighbourhood level social deprivation and the risk of psychotic
  disorders: a systematic review}.
\newblock \emph{Social psychiatry and psychiatric epidemiology}, 51:\penalty0
  941--950, 2016.

\bibitem[Penfold and Zhang(2013)]{penfold2013use}
R.~B. Penfold and F.~Zhang.
\newblock Use of interrupted time series analysis in evaluating health care
  quality improvements.
\newblock \emph{Academic pediatrics}, 13\penalty0 (6):\penalty0 S38--S44, 2013.

\bibitem[Rose et~al.(2020)Rose, Manning, Bentall, Bhui, Burgess, Carr, Cornish,
  Devakumar, Dowd, Ecks, et~al.]{rose2020social}
N.~Rose, N.~Manning, R.~Bentall, K.~Bhui, R.~Burgess, S.~Carr, F.~Cornish,
  D.~Devakumar, J.~B. Dowd, S.~Ecks, et~al.
\newblock The social underpinnings of mental distress in the time of
  covid-19--time for urgent action.
\newblock \emph{Wellcome open research}, 5, 2020.

\bibitem[Rue et~al.(2009)Rue, Martino, and Chopin]{rue2009approximate}
H.~Rue, S.~Martino, and N.~Chopin.
\newblock {Approximate Bayesian inference for latent Gaussian models by using
  integrated nested Laplace approximations}.
\newblock \emph{{Journal of the Royal Statistical Society: Series B
  (Statistical Methodology)}}, 71\penalty0 (2):\penalty0 319--392, 2009.

\bibitem[Singh et~al.(2019)Singh, Daniel, Baker, and Bentley]{singh2019housing}
A.~Singh, L.~Daniel, E.~Baker, and R.~Bentley.
\newblock Housing disadvantage and poor mental health: a systematic review.
\newblock \emph{American journal of preventive medicine}, 57\penalty0
  (2):\penalty0 262--272, 2019.

\bibitem[{United Kingdom Parliment}(2020)]{ukparliment2020aims}
{United Kingdom Parliment}.
\newblock \emph{The aims of ten years of welfare reform (2010-2020)}.
\newblock
  \url{https://commonslibrary.parliament.uk/research-briefings/cbp-9090/},
  2020.
\newblock {Accessed May 2023}.

\bibitem[{University of Essex, Institute for Social and Economic
  Research}(2019)]{ukhls2019weights}
{University of Essex, Institute for Social and Economic Research}.
\newblock \emph{Weighting and Sample Representation: Frequently Asked
  Questions}.
\newblock
  \url{https://www.understandingsociety.ac.uk/sites/default/files/downloads/general/weighting_faqs.pdf},
  2019.
\newblock {Accessed December 2022}.

\bibitem[{University of Essex, Institute for Social and Economic
  Research}(2022a)]{ukhls2022data}
{University of Essex, Institute for Social and Economic Research}.
\newblock \emph{Understanding Society: Waves 1-12, 2009-2021 and Harmonised
  BHPS: Waves 1-18, 1991-2009. 17th Edition. UK Data Service.}
\newblock \url{https://www.understandingsociety.ac.uk/}, 2022a.
\newblock {Accessed December 2022}.

\bibitem[Wagner et~al.(2002)Wagner, Soumerai, Zhang, and
  Ross-Degnan]{wagner2002segmented}
A.~K. Wagner, S.~B. Soumerai, F.~Zhang, and D.~Ross-Degnan.
\newblock Segmented regression analysis of interrupted time series studies in
  medication use research.
\newblock \emph{Journal of clinical pharmacy and therapeutics}, 27\penalty0
  (4):\penalty0 299--309, 2002.

\bibitem[Wickham et~al.(2020)Wickham, Bentley, Rose, Whitehead,
  Taylor-Robinson, and Barr]{wickham2020effects}
S.~Wickham, L.~Bentley, T.~Rose, M.~Whitehead, D.~Taylor-Robinson, and B.~Barr.
\newblock {Effects on mental health of a UK welfare reform, Universal Credit: a
  longitudinal controlled study}.
\newblock \emph{{The Lancet Public Health}}, 5\penalty0 (3):\penalty0
  e157--e164, 2020.

\bibitem[Zuccotti and O’Reilly(2019)]{zuccotti2019ethnicity}
C.~V. Zuccotti and J.~O’Reilly.
\newblock Ethnicity, gender and household effects on becoming neet: An
  intersectional analysis.
\newblock \emph{Work, employment and society}, 33\penalty0 (3):\penalty0
  351--373, 2019.

\end{thebibliography}

\clearpage

\end{document}